\documentstyle[aps,prb,twocolumn]{revtex}

\begin{document}

\title{Entropy Production via Particle
Production\thanks{to appear in: Proc. NATO Adv. Research
Workshop `Hot Hadronic Matter: Theory and Experiment,'
Divonne-les-Bains, June 27 -- July 1, 1994,
J. Rafelski (ed.)}}

\author{J. Rau}
\address{Max-Planck-Institut f\"ur Kernphysik,
Postfach 103980, 69029 Heidelberg,
Germany}

\date{October 22, 1994}

\maketitle

\begin{abstract}

I critically examine the notion of ``irreversibility,'' and
discuss in what sense it applies to the spontaneous creation
of particles in external fields.
The investigation reveals that particle
creation in very strong fields can only be described by a
non-Markovian transport theory.

\end{abstract}

\section{What is the Problem?}

\paragraph*{Spontaneous pair creation contributes to thermalisation.}
A heavy-ion collision appears highly irreversible:
while the initial state of the two colliding nuclei is almost pure,
particles emerging from the reaction can well be described by
a thermal distribution.
Some of the mechanisms responsible for this
thermalisation are well understood:
binary collisions of the microscopic constituents (partons),
hadrochemical reactions, or radiation.
At very high energies, however, an increasingly important role is played by
yet another elementary process: the spontaneous creation of
$q\overline q$-pairs in strong, coherent chromoelectric fields
(``flux tubes'').\cite{string}
These pairs are produced with a certain momentum distribution,
associated with which is a non-zero entropy.
It seems therefore that not only particles, but also entropy
is produced `out of the vacuum.'
This poses the following question:
Is spontaneous pair creation indeed irreversible?
And if so, in what sense?
\section{What is Irreversibility?}
\paragraph*{All investigations of ``irreversibility'' are based
upon a classification of the degrees of freedom.}
The v.Neumann entropy associated with the full statistical operator
of a closed quantum system,
\begin{equation}
S[\rho(t)]:=-k\,\mbox{tr}(\rho(t)\ln\rho(t))
\quad,
\end{equation}
is constant in time and hence for our purposes useless.
Non-trivial statements about irreversible
behavior all refer to a
time-dependent {\em relevant entropy},
which is the entropy associated with
the {\em relevant} degrees of freedom.
A distinction between relevant and irrelevant degrees of
freedom is thus essential.
Often, such a distinction appears in disguise:
as a classification of
observed vs. unobserved; observable vs. unobservable;
subsystem vs. environment; collective vs. noncollective;
slow vs. fast; shape vs. randomizing; or extrinsic vs. intrinsic
degrees of freedom.
\paragraph*{The relevant entropy is obtained by
coarse-graining.}\cite{jaynes}
The v.Neumann entropy measures the amount of
missing information as to the pure state of the system,
if one is given the expectation values
of {\em all} observables of the system.
In contrast, the relevant entropy measures the amount of
missing information as to the pure state of the system,
if one is given the expectation values
of only the {\em relevant} observables.
Going from the former to the latter involves discarding
information about
the irrelevant degrees of freedom -- a truncation
which is reflected in the
general inequality
\begin{equation}
S_{\rm rel}(t)\ge S[\rho(t)]\quad\forall\,\,t
\quad,
\end{equation}
and often
referred to as
``coarse-graining.''
\paragraph*{Example: Gibbs vs. Boltzmann entropy.}
The full state of a classical gas of $N$ indistinguishable
particles is given by a symmetric probability density
$W(\pi_1\ldots\pi_N)$ on the $6N$-dimensional $N$-particle
phase space. Associated with this full state is the
Gibbs entropy
\begin{equation}
S_G:=-k\int\mbox{d}^6\pi_1\ldots\mbox{d}^6\pi_N\,
W(\pi_1\ldots\pi_N)\ln W(\pi_1\ldots\pi_N)
\end{equation}
which, due to Liouville's theorem, stays constant
-- just like the v.Neumann entropy.
For reasons to be discussed shortly, one generally
considers only single-particle observables to be relevant,
but many-particle correlations to be irrelevant.
All information about the relevant degrees of freedom
is then contained in the
reduced probability density
\begin{equation}
w(\pi):=\int\mbox{d}^6\pi_2\ldots\mbox{d}^6\pi_N\,
W(\pi_1\ldots\pi_N)
\quad,
\end{equation}
defined on the 6-dimensional single-particle phase space.
Associated with this reduced state is the Boltzmann entropy
\begin{equation}
S_B:=-kN\int\mbox{d}^6\pi\,w(\pi)\ln w(\pi)
\quad.
\end{equation}
The fact that information about particle correlations has been
discarded is reflected in the inequality
\begin{equation}
S_B\ge S_G
\quad.
\end{equation}
It is the coarse-grained Boltzmann entropy to which
all non-trivial statements refer, including the famous $H$-theorem.
\paragraph*{The $H$-theorem relies on a strong separation of
time scales.}
A prediction cannot possibly contain more information
than the data on which it is based.
Therefore, if the evolution of
the relevant degrees of freedom is Markovian, i. e.,
if their expectation values
at time $t+\mbox{d}t$
can be predicted on the basis of their expectation values
at time $t$,
then the associated relevant entropy can only increase
or stay constant, but never decrease.
This is the $H$-theorem.
Clearly, its validity depends crucially
on the Markovian property of the evolution.
Consider, as an example, the classical gas:
If the gas is dilute, and hence the duration of one individual
collision much shorter than the average time that elapses
between two subsequent collisions,
then collisions may be regarded as statistically independent
(``Sto{\ss}zahlenansatz'');
the evolution equation (Boltzmann equation) becomes Markovian;
and therefore the $H$-theorem holds.
If, on the other hand, the gas is dense, and hence
the duration of a collision approximately equal to
the time between two collisions,
then there may be memory effects;
the evolution is no longer Markovian;
and the $H$-theorem may
be temporarily violated.\cite{daniel}
\paragraph*{``Irreversible'' is the flow of information from
slow to fast degrees of freedom.}
While the constancy of the v.~Neumann entropy shows
that complete information about the system is retained
in a full microscopic description,
the variation of the relevant entropy indicates
that the amount of information carried by the
relevant degrees of freedom continuously changes.
An obvious interpretation is that in the course of the
system's evolution,
information about the system
is being transferred between relevant and irrelevant
degrees of freedom.
How exactly the relevant entropy behaves, depends crucially on the
choice of the ``relevant vs. irrelevant'' classification.
In general,
the flow of information need not have a unique direction;
it may occur either from relevant to irrelevant, or
from irrelevant to relevant degrees of freedom.
As a consequence, the relevant entropy may either increase
or decrease.
It is
only when the relevant degrees of freedom are slow and the
irrelevant degrees of freedom are fast that, according to the
general $H$-theorem, the information flow
is uniquely directed from relevant to irrelevant
degrees of freedom.
This ``leaking'' of information into fast degrees of freedom
is perceived as irreversible.
However, the information is not really lost -- it only becomes
inaccessible to a certain coarse-grained level of description.
In the example of a dilute classical gas,
information is being transferred
from single-particle observables to many-particle
correlations.
Since in general these correlations will not be measured,
part of the information about
the system becomes experimentally inaccessible.
\paragraph*{The study of irreversible behavior reduces to a
time scale analysis.}
``Irreversibility''
refers to the experimenter's ability -- or the lack thereof --
to prepare, control,
or monitor certain degrees of freedom;
it thus seems to be
a purely ``anthropomorphic'' concept.
Nevertheless, irreversible features of the dynamics are
not entirely subjective.
For an observable to be measurable in practice, it is
usually necessary that it varies slowly.
In this case the experimentally monitored degrees of freedom constitute
some subset of the set of all {\em slow} degrees of freedom.
Accordingly, the flow of information from observed to
unobserved degrees of freedom is intimately tied to the
flow of information from slow to fast degrees of freedom.
But the latter is an {\em objective} property of the dynamics: it is
determined by the presence of disparate time scales, and not
dependent on any observer.
A study of irreversible features of the dynamics should
therefore focus on the identification of slow and fast degrees
of freedom, and on the analysis of the associated time scales.
The choice of the ``relevant vs. irrelevant'' classification
is then no longer as arbitrary and subjective as it may have seemed
at first;
rather, the proper choice is determined by objective physical criteria
-- by the time scales.
\section{Is Spontaneous Pair Creation Irreversible?}
\paragraph*{The relevant degrees of freedom are the occupation
numbers of momentum states.}
What is commonly being measured in an experiment, and what
enters into most transport equations such as the quantum
Boltzmann equation, are the momentum distributions
$n_\mp(\vec p,t)$ of the produced
particles and antiparticles.
This strongly suggests choosing the occupation numbers of
momentum states as the relevant degrees of freedom.
Associated with this choice is the relevant entropy
(for spin-$1/2$ fermions)
\begin{eqnarray}
S_{\rm rel}(t)&:=&
-2k\int_{\vec p}\left[{n_-\over2}\ln{n_-\over2}+
\left(1-{n_-\over2}\right)\ln\left(1-{n_-\over2}\right)\right.
\nonumber \\
&&
\mbox{} +\left.
\,(n_-\leftrightarrow n_+)\,\right]
\label{rel_ent}
\end{eqnarray}
(where $n_\mp\equiv n_\mp(\vec p,t)$).
Whether or not the relevant
degrees of freedom are sufficiently slow; i. e.,
whether or not their evolution is Markovian;
whether or not the relevant entropy obeys an $H$-theorem;
and hence whether or not in this description pair
creation appears irreversible --
all this must be the subject of a thorough time scale analysis.
\paragraph*{I focus on the pair creation process proper.}
Irreversible information flows occur at two different stages
of the evolution: (i) during the pair creation process proper,
and (ii) during the subsequent ``decoherence.''
During the pair creation process proper, information leaks
from relevant occupation numbers into correlations and
relative phases.
Under the influence of subsequent collisions, this
phase information is then being transferred further to an
unobservable ``environment''
(or ``heat bath'') of high-frequency partons.
Such a transfer of phase information to an unobservable
environment is often referred to as ``decoherence,'' and discussed
elsewhere in these Proceedings.\cite{elze}
Here, I wish to concentrate on stage one of the evolution,
the pair creation process proper.
\paragraph*{A time scale analysis requires knowledge
of the non-Markovian equation of motion.}
The question to be addressed is the following:
Is the initial flow of information from occupation numbers to
correlations and relative phases irreversible?
In other words,
does the relevant entropy (\ref{rel_ent}) obey an $H$-theorem?
According to our previous discussion, the answer to this question
hinges upon a careful analysis of time scales.
It is necessary to derive a -- generally non-Markovian! --
equation of motion for the occupation numbers;
to identify the memory time of this non-Markovian equation
as well as the time scale on which the occupation numbers
evolve; to compare these, and thus to find a criterion for
the validity of the Markovian approximation.
If and only if this criterion is satisfied, the
$H$-theorem holds. If, on the other hand, the criterion is {\em not}
satisfied, then the memory time furnishes the
typical time scale on which the relevant entropy
may temporarily decrease.
\paragraph*{Microscopic model: Schwinger mechanism.}\cite{schwinger}
All essential features of the pair creation process are
exhibited by the well-known Schwinger mechanism,
the spontaneous creation of $e^+e^-$ pairs
in a constant, homogeneous external electric field.
It is this simple model which has been considered in much of the
literature,\cite{string} and which I now want to subject to
further analysis.
Let $q$ be the electron charge, $\vec E$ the external field,
$\vec p(t):=\vec p+q\vec E t$ the time-dependent momentum,
$m$ the spin component, and $\phi_{fi}$ the dynamical phase
accumulated between times $t_i$ and $t_f$.
Then in the Heisenberg picture the evolution mixes
particle ($a^\dagger$) and antiparticle ($b$) field operators,
with respective amplitudes $\alpha_{fi}$ and $\beta_{fi}$:
\begin{eqnarray}
\lefteqn{{\cal U}(t_2,t_1)
\left(
\begin{array}{c}
a^\dagger(\vec p(t_1),m)\\
b(-\vec p(t_1),-m)
\end{array}
\right)
=}
\nonumber \\
&& \left(
\begin{array}{cc}
\alpha_{21} & \beta_{21} \\
-\beta_{21}^* & \alpha_{21}^*
\end{array}
\right)
\left(
\begin{array}{cc}
\mbox{e}^{{\rm i}\phi_{21}} & 0 \\
0 & \mbox{e}^{-{\rm i}\phi_{21}}
\end{array}
\right)
\left(
\begin{array}{c}
a^\dagger(\vec p(t_2),m)\\
b(-\vec p(t_2),-m)
\end{array}
\right)
\end{eqnarray}
Hence on the microscopic level, spontaneous pair creation is
described by a time-dependent Bogoliubov transformation.
\paragraph*{In the equation of motion, pair creation is accounted
for by a non-Markovian source term.}
The equation of motion for the occupation numbers has
the structure of a quantum Boltzmann equation.
Aside from the usual acceleration and
(possibly) collision terms, it contains an additional source term
to account for the spontaneous pair creation.
By means of the so-called projection method,\cite{projection}
this source term may be related to the coefficients of the
Bogoliubov transformation:\cite{rau_prd}
\begin{eqnarray}
\dot n_-^{\rm sou}(\vec p,t)
&=&
4\,{\rm Re}\int_0^{t-t_0}{\rm d}\tau\,
\dot\beta^*(-\tau,-\tau)\,{\rm e}^{-i2\phi(-\tau,0)}
\dot\beta(0,0)
\nonumber \\
&&\times S(\vec p-q\vec E \tau,t-\tau)
\end{eqnarray}
Here $t_0$ denotes the initial time at which the
external field is switched on;
$\dot\beta(t_1,t_1)$ is a shorthand for $\partial\beta(t_2,t_1)/
\partial t_2|_{t_2=t_1}$; and the factor $S\equiv
[1-n_-/2][1-n_+/2]-(1/4)n_-n_+$ accounts for Pauli blocking
as well as the possible annihilation of pairs back into the field.
Evidently, the source term
involves an integration over the entire history of the
system and is therefore non-Markovian.
\paragraph*{The memory time combines a quantum mechanical and
a classical time scale.}
Careful analysis\cite{rau_prd} reveals that significant
contributions to the above source term come only from times $\tau$
which are smaller than the characteristic memory time
\begin{equation}
\tau_{\rm mem}\sim {\hbar\over\epsilon_\perp}+
{\epsilon_\perp\over qE}
\end{equation}
(where $\epsilon_\perp$ denotes the transverse energy or mass).
The two terms have very different physical origins.
(i) The time $\hbar/\epsilon_\perp$ is quantum mechanical.
It corresponds -- via the time-energy uncertainty relation --
to the time needed to create a virtual particle-antiparticle
pair, and may thus be regarded as the ``time between two
production attempts.''
(ii) The time $\epsilon_\perp/qE$, on the other hand, is classical.
It can be interpreted in various ways, depending on the picture
employed to visualize the pair creation process.
If pair creation is viewed as a tunneling process from the
negative to the positive energy continuum, this classical time
coincides with the time needed for the wave function to traverse
the barrier with the speed of light.
\paragraph*{The Markovian approximation is valid only for
weak fields.}
The occupation numbers evolve
on a  typical time scale set by
the inverse production rate $[\dot n_-^{\rm sou}(\vec p\,)]^{-1}$.
Assuming $p_{\|}=0$ for simplicity, one obtains as a
representative scale
\begin{equation}
\tau_{\rm prod}\sim
{\epsilon_\perp\over qE}\exp\left({\pi\epsilon_\perp^2
\over 2\hbar qE}\right)
\quad.
\end{equation}
This production scale is much larger than the memory time,
and hence the evolution is Markovian,
only if $E\ll m^2/\hbar q$.
\paragraph*{First conclusion: The flow of information from
occupation numbers to correlations and phases is strictly
irreversible only if the field is weak.}
As long as the field is sufficiently weak, the evolution of
occupation numbers is Markovian, the $H$-theorem holds, and
hence the relevant entropy increases monotonically.
But as soon as $E\sim m^2/\hbar q$, both memory time and production
scale attain the same magnitude $\tau\sim\hbar/m$.
As a result, the Markovian approximation breaks down, and there may be
temporary violations of the $H$-theorem (on the same time scale
$\tau\sim\hbar/m$).
Indeed, oscillations of the relevant entropy have been observed
in numerical simulations.\cite{cooper}
\paragraph*{Second conclusion: Pair creation
in very strong fields must be described by a non-Markovian
transport theory.}
Whenever the external field is stronger than the critical
value $m^2/\hbar q$, there may be sizeable memory effects
which cannot be accounted for in a Markovian
transport theory.
$[$Particle masses of the order 0.5, 15 or 500 MeV correspond
to critical field strengths of the order
$10^{-3}$, 1 or $10^3$ MeV/fm, respectively.$]$
This strong-field domain is in fact the only one which is physically relevant,
because only there does spontaneous pair creation
occur at an appreciable rate.
Hence for all practical purposes pair creation must {\em always} be
described by a {\em non}-Markovian transport theory.

\acknowledgements
I thank B. M\"uller for advice and many useful discussions.
Financial support by the Studienstiftung des deutschen Volkes,
the U.S. Department of Energy (Grant DE-FG05-90ER40592) and the Heidelberger
Akademie der Wissenschaften
is gratefully acknowledged.

\vfill

\end{document}